\documentstyle[preprint,eqsecnum,aps,prb]{revtex}

\begin{document}
\draft
\title{{\em ab initio} statistical mechanics of the ferroelectric 
phase 
transition in $PbTiO_3$}
\author{U. V. Waghmare\cite{um} and K. M. Rabe}
\address{Dept of Applied Physics, Yale University \\
P. O. Box 208284, New Haven, Connecticut, 06520-8284}
\date{\today}
\maketitle
 
\begin{abstract}
 
\noindent
An effective Hamiltonian for the ferroelectric transition in 
$PbTiO_3$ is constructed from first-principles 
density-functional-theory total-energy and
linear-response calculations through the use of a localized, 
symmetrized basis
set of ``lattice Wannier functions.'' 
Explicit parametrization of the polar lattice Wannier functions is
 used
for subspace projection, addressing the issues of LO-TO splitting 
and coupling to the complementary subspace.
In contrast with ferroelectric $BaTiO_3$ and $KNbO_3$, we find 
significant
involvement of the $Pb$ atom in the lattice instability.
Monte Carlo simulations for this Hamiltonian show a
first-order cubic-tetragonal transition at 660 K. Resulting 
temperature 
dependence of spontaneous polarization, $c/a$ ratio and unit-cell 
volume near 
the transition are in good agreement with experiment.
Comparison of Monte Carlo results with mean field theory analysis
shows that both strain and fluctuations are necessary to
produce the first order character of this transition.
\end{abstract}

 
\narrowtext
 
\section{INTRODUCTION}

Perovskite-structure oxides exhibit various types of lattice 
instabilities
resulting from inherent structural frustration in the prototype cubic 
structure
\cite{lg}, shown in Fig. 1(a).
This class of materials includes a large number of ferroelectrics, 
with uniform polar distortions and
accompanying lattice relaxation (eg. $PbTiO_3, BaTiO_3, KNbO_3$), 
while cation
substitution can result in dramatic changes in ground state distortion
(eg. antiferroelectric in $PbZrO_3$, antiferrodistortive in $SrTiO_3$) 
and corresponding complexities in the mixed systems (eg. 
$PbZr_{1-x}Ti_xO_3,
Ba_{1-x}Sr_xTiO_3$). However, in nearly all examples,
the amplitudes and energies of the distortions are
rather small, and cubic symmetry is restored at temperatures above a 
critical temperature $T_c$, typically a few hundred degrees Kelvin.

For a better understanding of structural phase transitions in 
perovskite
oxides, including chemical trends in the transition temperatures,
the first-order vs second-order character of transitions,
the relationship between local distortions and average 
crystallographic
structure, and the stability of intermediate temperature phases,
first-principles calculations offer valuable access to microscopic 
information.
With advances in algorithms and computational capabilities,
the challenge of achieving the
high accuracy necessary for studying these distortions has been 
largely met,
and ground state distortions well reproduced for a wide range of
perovskite-structure oxides\cite{ferroc,ksv,ferros}.
However, for ab initio molecular dynamics or Monte Carlo, the system
sizes required for the study of finite-temperature structural 
transitions are still completely impractical.

An alternative approach is to focus on a restricted subset of the
degrees of freedom that is relevant to the transition and construct
a simple effective Hamiltonian in this subspace.
The parameters in these models are chosen to reproduce the low energy 
surface of an individual material, and thus to reproduce its finite 
temperature behaviour in the vicinity of its transition.
Comparison of these models gives a systematic
understanding of similarities and differences in the microscopic 
structural
energetics of
different materials. From the dependence of calculated properties on 
effective Hamiltonian parameters, one can also obtain a better 
understanding
of the role of various microscopic couplings in producing the 
observed 
behaviour. Microscopic effective Hamiltonians for structural 
transitions
in perovskites were first constructed using the concept of local 
modes,
with empirically determined parameters.\cite{pytte,lines}
First-principles total energy calculations were used in the 
determination
of a local mode effective Hamiltonian for the structural transition 
in
$GeTe$\cite{rj} and more recently for the structural transitions in 
$PbTiO_3$\cite{rw2}, $BaTiO_3$\cite{zvr} and $SrTiO_3$\cite{zvsto}.
A systematic approach which generalizes and refines the local mode 
concept,
allowing the efficient construction of an optimal effective 
Hamiltonian
from first-principles total-energy and linear response techniques,
has been developed based on the concept of lattice Wannier 
functions.\cite{rwwannier}
This approach exploits symmetry properties of the system and is 
generally
applicable to complex structural transitions involving several 
unstable 
modes
including ones at the zone boundary.
Information from additional first principles calculations allows 
for a 
systematic check on the validity of the truncation of the effective 
Hamiltonian,
and, when needed, the expansion of the subspace and refinements of 
its form.
The resulting effective Hamiltonian
is quantitatively realistic while retaining a simple and physically
transparent form.

In this paper, we present a detailed description of the
first-principles investigation of $PbTiO_3$\cite{rwjpcs}, which 
exhibits a single 
first-order transition at 763 K from the cubic high-temperature phase 
to the 
ferroelectric tetragonal ground state, shown in Fig. 1(b).
We construct an effective Hamiltonian for this structural 
phase transition from first principles using the lattice Wannier 
function method. 
In contrast with $BaTiO_3$\cite{zvr} and $KNbO_3$\cite{yuk} for which
 the 
uniform polar distortions in the low temperature phase consist of 
predominantly
B-atom displacements, those in $PbTiO_3$ are dominated by A-atom 
($Pb$)
displacements, which will be important in determining the effective 
Hamiltonian subspace.
The effective Hamiltonian also contains the coupling of these local 
polar distortions to strain.
In Ref. \onlinecite{ferroc}, the tetragonal phase in $PbTiO_3$ was
 found to be
stabilized relative to the rhombohedral phase by the unit cell 
relaxation.  In addition, we will find that 
strain plays a crucial role in producing the correct finite 
temperature transition behaviour.

In Section IIA, we briefly review the method of lattice Wannier 
functions.
In Section IIB and IIC, the first principles methods and results 
obtained for 
the lattice constant, elastic constants, phonon frequencies and the 
effective charges of $PbTiO_3$ are presented.
In Sections IIIA and IIIB, we describe the construction of
the effective Hamiltonian, with particular attention to the
treatment of LO-TO splitting and crossing of branches through 
explicit parameterization of the lattice Wannier functions.
In Section IIIC, we describe properties of the ground state of the
 effective Hamiltonian determined from first principles.
In Section IV, we present results of finite temperature analysis of 
$H_{eff}$ using mean field theory and Monte Carlo simulations.
These results are discussed in Section V.

\section{Method}

\subsection{Lattice Wannier Function method for the construction 
of $H_{eff}$}

In the lattice Wannier function method, the effective Hamiltonian is
 obtained
as the result of projection of the full lattice Hamiltonian (in the
Born-Oppenheimer approximation) into a subspace of the full ionic 
displacement space.  The effective Hamiltonian subspace is spanned by 
an orthonormal basis of ``lattice Wannier functions:'' symmetrized 
localized
atomic displacement patterns taken with respect to a high-symmetry 
reference configuration.
This basis defines a set of coordinates such that a given set of 
values of the 
coordinates corresponds directly to a particular pattern of atomic
displacements. 
As a result of the symmetrized and localized nature of the basis, the 
Taylor
expansion of the effective Hamiltonian around the high-symmetry 
reference
configuration (with all coordinate values equal to zero) has a simple 
form with
relatively few parameters, which can be determined from first 
principles
calculations using the correspondence to patterns of atomic  
displacements.

We briefly review the procedure; further details can be found in 
Ref. \onlinecite{rwwannier}.
Construction of the subspace begins with a Taylor expansion of the full
lattice Hamiltonian to quadratic order. A subset of the eigenvectors of
the quadratic Hamiltonian is selected for inclusion in the subspace.
 This 
subset must include the unstable modes which freeze in to produce the
low-temperature structure. In addition, to achieve a good description
 of the
branches which emanate from the unstable modes, ``end-points" of these
branches at high symmetry k-points are included.
The symmetry properties of the subspace are established by identifying 
symmetries of localized functions (Wyckoff position and site symmetry 
group irrep) which can build up the selected subset of modes. 

We follow the prescription in Ref. \onlinecite{rwwannier} to
obtain an explicit, though approximate, form of a 
lattice Wannier basis vector. This involves finding the symmetric
coordination shells surrounding a representative Wyckoff site and 
identifying the independent
displacement patterns of each shell that transform according to the 
given irreducible representation of the site symmetry group. 
The amplitudes of
these displacement patterns completely specify an LWF. 
Because of the localized nature of LWFs, this infinite number of
parameters can, to a good approximation, be reduced to a small finite 
number
by neglecting the displacements of shells beyond a chosen range.
At high symmetry points in the BZ, the modes built up with these 
parametrized LWFs are then fit to the 
corresponding normalized mode eigenvectors obtained from first 
principles.

These basis functions completely specify the effective Hamiltonian 
subspace. In the ideal case, this subspace is completely decoupled at 
quadratic order
from its complementary subspace and to a good approximation at higher 
order as well.
This happens when the subspace consists of entire branches isolated in 
energy from the
others and contains all the unstable modes. In most real systems, 
branches emanating 
from the unstable modes cross with branches in the complementary 
subspace. 
This leads to some degree of quadratic coupling which is unimportant 
if the crossing occurs
far away from the unstable modes. If not, the subspace should be 
expanded to 
include these
branches. In addition, in polar crystals, the electric field at 
$\vec q \rightarrow 0$ can mix
the LO modes differently from the corresponding TO modes. In such a
 case, 
the Wannier basis vector which reproduces a given TO branch will not 
reproduce any LO mode exactly. However,
since LO modes are typically high in energy, this approximate 
description of 
the LO mode
should be adequate for the description of the structural transition.

The quadratic part of the lattice Hamiltonian has both kinetic and 
potential 
energy contributions.
However, in the classical statistical mechanical treatment, the 
kinetic energy terms appear 
in Gaussian integrals in the partition function, factoring
out to give a trivial contribution to the free energy.
In this case, the eigenmodes used in the construction above could 
be of either the force constant matrix or
the dynamical matrix. In $PbTiO_3$, we have found that the difference
 in the resulting effective Hamiltonian subspace is rather small and
both choices should give comparable results.
In the construction described in section III, we have used the 
eigenmodes of the force constant matrix.
Eigenmodes of the dynamical matrix  
are strongly preferable only if the effective Hamiltonian is also 
to be used in 
classical dynamics or quantum mechanical simulations, 
since in that case the form of the kinetic energy is greatly 
simplified.

In the final step, the lattice Hamiltonian is projected into this 
subspace to obtain the effective Hamiltonian.
An explicit form $H_{eff}$ is obtained by identifying a small number 
of physically important terms in a Taylor expansion in the lattice 
Wannier coordinates.
The coefficients of these terms are parameters to be determined 
from first principles by fitting $H_{eff}$ to the results
of selected total energy and linear response calculations, using 
the explicit
correspondence between the Wannier coordinates and the
actual ionic displacements. To check the validity of the truncated 
form of the effective Hamiltonian, additional independent first 
principles calculations 
can be performed and compared with $H_{eff}$.

\subsection{First principles total energy calculations}

The first-principles calculations for $PbTiO_3$ were
performed using the ab initio pseudopotential method
in the local density approximation (LDA) with the Perdew-Zunger 
parametrization of the Ceperley-Alder density functional.\cite{cap}
For Pb, the scalar-relativistic pseudopotentials from Ref. 
\onlinecite{bhs} were used.
The use of a plane wave basis set dictates the use of optimized
pseudopotentials\cite{opt}
for O and Ti to achieve reasonable energy convergence and 
transferability.
For O, the reference configuration $2s^22p^4$ was used with 
pseudopotential core radii $r_{c,s}=r_{c,p}=1.5~a.u.$
Optimization was performed with $q_{c,s}=7.0(Ry)^{1/2}$ and 
$q_{c,p}=6.5(Ry)^{1/2}$ and 4 and 3 Bessel
functions for s and p orbitals respectively.
For Ti, it is essential to treat the semi-core $3s$ and $3p$
electrons as valence electrons\cite{ksv,ghosez}. 
The reference configuration $3s^23p^63d^2$ was
used with pseudopotential core radii $r_{c,s}=r_{c,p}=1.45~a.u.$ and
$r_{c,d}=1.5~a.u.$. Optimization was performed with 
$q_{c,s}=7.2(Ry)^{1/2}$, $q_{c,p}=7.0(Ry)^{1/2}$ and
$q_{c,d}=7.74(Ry)^{1/2}$, and 4 Bessel functions.
An energy cutoff of $850$ eV (corresponding to approximately 3600 
plane waves 
for a 5-atom unit cell) 
was used to ensure convergence within $10 mRy$/atom.
The self-consistent total energy calculations were performed using the
program CASTEP 2.1\cite{castep}
based on the stable and efficient preconditioned conjugate-gradients 
method \cite{payne}.
For the Brillouin zone integrations, k-point sampling was performed 
using the Monkhorst-Pack construction\cite{mp} with
64 k-points in the full Brillouin zone.

As reported in Ref. \onlinecite{rw2} and summarized here in Table (I), 
the lattice constant and elastic constants
of $PbTiO_3$ in the cubic perovskite structure obtained from the total 
energy calculations for a range of unit cell volumes are in good
agreement with previous calculations\cite{ferroc,ksv}.
In addition, in Fig. 4 of Ref. \onlinecite{rw2} we showed the 
calculated 
energies as a function of experimental soft mode amplitude, 
which compare favorably with previous LAPW 
calculations\cite{ferroc}.

\subsection{First principles DFT linear response}

The technique of DFT linear response is used to calculate the second
derivatives of the total energy with respect to perturbation 
parameters
through the self consistent calculation of the first order correction
to the occupied Kohn-Sham wave functions\cite{bgt,gat}. For example, 
Born effective charges, 
dielectric constant and dynamical matrices are the second derivatives
of total energies and thus can be obtained with this technique. 
In this framework, the dielectric constant can be calculated avoiding
cumbersome sums over unoccupied bands.
Another significant advantage is that
$\vec q \neq 0$ force constants can be computed with an effort 
similar to that of a single unit cell total energy calculation.

Our implementation is a modification of CASTEP 2.1 based on the
variational formulation of DFT linear response\cite{gat}.
All the linear response calculations
reported here were done at the experimental lattice constant
\cite{shirane1} of $3.96883 \AA$
with 64 Monkhorst-Pack k-points in the full Brillouin zone. 
The Vosko-Wilk-Nusair parametrization of the Ceperly-Alder density 
functional
was used to permit the calculation of derivatives of 
exchange-correlation 
terms\cite{vosko}.
A $36 \times 36 \times 36$ fourier transform grid is used
for integration over a unit cell in real space. 
This real space grid breaks global
translational invariance\cite{ggps,gat}, which manifests itself
as small violations of the acoustic sum rule (the calculated 
frequencies of 
zone center acoustic modes are not exactly zero) and charge neutrality
(the calculated change in polarization due to a rigid displacement
 of the crystal in any direction is not exactly zero). 
The acoustic sum rule was imposed by adding small corrections to the
diagonal elements of the $\vec q = 0$ force constant matrix.
Charge neutrality was imposed by 
adding the same small correction to the effective charges of all 
atoms.

The Born effective charges are presented in Table (II) in very good
 agreement
with previously calculated values\cite{zkv}. The main features of 
interest are
the anomalously large effective charges of Ti and O along the bond 
and the anisotropy of the oxygen charge.
The calculated dielectric constant is 8.24, which can be compared with 
the experimental value of 8.64 quoted in Ref. \onlinecite{zkv}.
The data in Table (II) combined with the calculated force constants 
at $\vec q = 0$ give 
the frequencies of IR-active phonons presented in Table (III).
Direct comparison of these results with the previous 
calculations\cite{zkv}
is not possible because the calculations were performed at different 
lattice constants.
This has an especially large impact on the unstable mode frequency,
as confirmed by our calculations of coupling between this mode 
and homogeneous strain, to be described below. 
As can be seen in Table (IV), the unstable $\Gamma_{15}$ mode
has the largest mode effective charge, which should be associated 
with the largest LO-TO splitting. 
Since there are three polar zone center modes with the same symmetry,
mixing leads to LO-mode eigenvectors different from 
TO-mode eigenvectors.
Effects of this mixing can be quantified using the correlation 
matrix\cite{zkv}
\begin{equation}
c_{ij}= <\xi_{i}^{TO}|M|\xi_{j}^{LO}>,
\end{equation}
given in table (IV),
where $M_{mn}=M_{m} \delta_{mn}$ is the mass matrix and $\xi_{i}$ 
are the IR-active mode eigenvectors. 
As expected, the unstable $\Gamma_{15}$
TO mode has the strongest correlation with the highest LO mode.

In Table V, we report selected phonon frequencies at other high 
symmetry
k-points in the Brillouin zone, focusing in particular on the lowest
energy phonons. 
While we also find unstable modes away from the $\Gamma$
point, the unstable mode at $\Gamma$ is clearly the dominant lattice
 instability
in our calculations, consistent with the observed low-temperature  
structure.
The eigenvectors of the lowest energy phonons and the corresponding
force constant matrices will be used in the determination of
the parameters in the effective Hamiltonian in the next section.

\section{Construction of the effective Hamiltonian for $PbTiO3$}

\subsection{Construction of the subspace}

The construction of the effective Hamiltonian subspace begins with 
consideration of the 
calculated force constant matrix eigenvalues and eigenvectors 
at $\Gamma, R, M$ and $X$. 
The subspace has to include the unstable polar $\Gamma_{15}$ mode 
which freezes in
to produce the low temperature tetragonal structure.
In addition, to achieve a good description of branches which  emanate 
from this
dominant unstable mode, the endpoints of these branches $R_{15}, 
M_2^{\prime}, M_5^{\prime}, X_5^{\prime}$ are included. 
As can be seen from Table (I) of Ref. \onlinecite{rwwannier}, 
the lattice Wannier functions which can build up this subset of modes 
transform like 3-dimensional vectors centered at Pb-sites.
It should be noted that the lowest mode at R is actually $R_{25}$,
which corresponds to an oxygen octahedron rotation instability  
seen in many perovskite oxides\cite{zvsto}. Since crossing of the
lowest branch along (111) with that emanating from $R_{25}$ 
occurs far from the relevant mode $\Gamma_{15}$
and relatively higher in energy, we do not include it in the subspace.

To include coupling of the relevant polar distortions ($\Gamma_{15}$)
 to
local distortions of the unit cell (inhomogeneous strain), we expand 
the 
subspace to include the acoustic modes 
by choosing an additional set of lattice Wannier functions.
Of the three possibilities (listed in Table (I) of Ref. 
\onlinecite{rwwannier}), 
Ti-centered 3-dimensional vectors are preferable to $O_{x,x}$ 
(1-dimensional 
vectors) and $O_{x,y}$ (2-dimensional vectors),
since this choice corresponds to the smallest subspace expansion and 
highest site symmetry group.
Furthermore, unlike $O_{x,x}$, the resulting $6\cdot N$ dimensional
subspace does not include the highest energy modes.

Next, we obtain an explicit form for the Pb-centered LWF.
This involves finding the symmetric
coordination shells surrounding a Pb site and
identifying the independent displacement patterns of each shell that 
transform 
according to the vector representation of the site symmetry group
 $O_h$.  For a given shell there can be more
than one pattern of displacements with a given transformation 
property. 
To each such pattern corresponds an independent amplitude parameter.
By neglecting the displacements of shells beyond 1st neighbour Ti and 
O shells and 2nd neighbour Pb shells,
we obtain a total of 10 parameters.
The first shell of Ti atoms has 2 independent displacement
patterns, parametrized by $t_1$ and $t_2$; there are 1, 2, 2 
parameters for 
the zeroth, first and second shells of Pb atoms respectively and 3
parameters for the first shell of oxygen atoms. These displacement 
patterns are
shown in Fig. 2.

To determine the numerical values of 
these parameters for $PbTiO_3$, we build up the transverse modes 
$e_{\vec q,\alpha}$ at high symmetry k-points in
the Brillouin zone, namely $\Gamma$, X, M, and R, from the 
parametrized LWF using 
\begin{equation}
e_{\vec q,\alpha} = \sum_{\vec R_{i}} exp(i\vec q \cdot \vec R_{i} )
w_{i,\alpha}
\end{equation}
where $\vec R_i$ is a direct lattice vector and $w_{i,\alpha}$ is an 
LWF 
centered at the Pb site in the {\it i}th unit cell, $\alpha$ being its 
Cartesian component.
This specifies atomic displacements in these modes as linear functions 
of the parameters, to be fit to the normalized eigenvectors of the
force constant matrix calculated from first principles.
With the parameters listed above, we can reproduce the normalized 
eigenvectors
of the modes $\Gamma_{15}$, $R_{15}$ and $M_2^\prime$ exactly.
The remaining free parameters, associated with Pb and Ti 
displacements, were used to fit to a normalized 
mode with maximum overlap with the lowest $M_5^\prime$ (see 
Table (VI)). 
Numerical values of these parameters, listed in Table (VII), clearly     
show that the magnitude of the parameters decays rapidly with 
shell-radius,
confirming the assumption of LWF localization. Furthermore,
by adding an additional shell of Pb (one parameter) and oxygen atoms
(two parameters), we can reproduce
all the transverse optical modes in the subspace. The parameter values 
for this refined LWF are given in Table (VII). The values of 
the parameters of the innermost shells do not change very much, 
and the values of the new parameters are very small.

Another way of testing the approximate LWF is to see how well it
reproduces other modes in the subspace.
For example, in Table VIII, we show the comparison of the first 
principles
$X_5^\prime$ eigenvector with the mode constructed with the 
approximate LWF.
The approximate mode has an overlap of $92 \%$ with the 
relevant mode, and if the approximate mode vector is normalized, the
overlap becomes $99.96 \%$, showing that the LWF describes the
subspace very well. 

For the simplest treatment of inhomogeneous strain (the acoustic 
branches),
an explicit expression of the Ti-centered LWF is not needed, since the
 goal is
only to reproduce the long wavelength acoustic modes, whose dispersion
 is determined from the elastic constants.
For a more refined treatment, an LWF
could be parametrized as above and 
determined by fitting to the first principles eigenmodes $\Gamma_{15}$,
$R_{25}^\prime$, $M_3^\prime$, $M_5^\prime$, $X_1$, $X_5$.

\subsection{Determination of parameters in $H_{eff}$}

Using the symmetry properties of the lattice Wannier basis for the 
effective  Hamiltonian subspace,
we write an explicit expression for $H_{eff}$ as a
Taylor expansion in the lattice Wannier coordinates, invariant under
 the space group $Pm3m$.
$\{\vec \xi_{i}\}$ and $\{\vec u_{i}\}$ denote the Pb-centered and
Ti-centered lattice Wannier coordinates respectively. 
Each of these three dimensional vector degrees of
freedom transforms according to the $\Gamma_{15}$ irrep
of the point symmetry group $O_h$. 
Below, we organize the terms in the expansion of $H_{eff}$ 
into those acting exclusively in the Pb-centered subspace and the 
Ti-centered subspace and those coupling the two.

In the Pb-centered subspace, 
we consider quadratic interactions 
up to third nearest neighbour with the most general form
allowed by the space group symmetry:
$$\sum_i(A|\vec \xi_i |^2$$
\begin{eqnarray}
\nonumber
+\sum_i\sum_{\hat d=nn1}[
a_L(\vec\xi_i\cdot \hat d)(\vec\xi_i(\hat d)\cdot \hat d)
+a_T(\vec\xi_i\cdot\vec\xi_i(\hat d)
-(\vec\xi_i\cdot \hat d)(\vec\xi_i(\hat d)\cdot \hat d))] \\
\nonumber
+\sum_i\sum_{\hat d=nn2}[
b_L(\vec\xi_i\cdot \hat d)(\vec\xi_i(\hat d)\cdot \hat d)
+b_{T1}(\vec\xi_i\cdot \hat d_1)(\vec\xi_i(\hat d)\cdot \hat d_1)
+b_{T2}(\vec\xi_i\cdot \hat d_2)(\vec\xi_i(\hat d)\cdot \hat d_2)]\\
+\sum_i\sum_{\hat d=nn3}[
c_L(\vec\xi_i\cdot \hat d)(\vec\xi_i(\hat d)\cdot \hat d)
+c_T(\vec\xi_i\cdot\vec\xi_i(\hat d)
-(\vec\xi_i\cdot \hat d)(\vec\xi_i(\hat d)\cdot \hat d))],
\end{eqnarray}
where $\vec\xi_i(\hat d)$ denotes the LWF coordinate at a neighbour 
of site i in $\hat d$ direction.
Beyond third neighbor we use a dipole-dipole form parametrized by the
mode effective charge $\overline Z^*$ and the electronic dielectric 
constant $\epsilon_\infty$:
\begin{equation}
\sum_i\sum_{\vec d}{(\overline Z^\star)^2 \over \epsilon_\infty}
{(\vec\xi_i\cdot\vec\xi_i(\hat d)-
3(\vec\xi_i \cdot \hat d)(\vec\xi_i(\hat d)\cdot \hat d))\over 
|\vec d|^{3}}.
\end{equation}
An important simplifying approximation is that the onsite potential,
depending on the value of $\xi_i$ at a single i, is the only set of
terms including anharmonic interactions acting exclusively in the 
Pb-centered subspace.
For simplicity, anharmonic terms are included only in the onsite 
potential with isotropic terms up to eighth order in $\vert \vec
\xi_i\vert$
and full cubic anisotropy at fourth order:
\begin{equation}
\sum_i(B|\vec \xi_i |^4+
C(\xi_{ix}^4+\xi_{iy}^4+\xi_{iz}^4)
+D|\vec \xi_i |^6+E|\vec \xi_i |^8).
\end{equation}

In the Pb-centered subspace, the parameters to be determined from 
first 
principles are $A, a_L, a_T, b_L, b_{T1}, b_{T2}, c_L, c_T, B, C, 
D, E, \overline Z^{\star}$.
This determination relies on the explicit correspondence 
between the lattice Wannier coordinate $\{\vec \xi_i\}$
and the ionic displacements $\{\vec d_i\}$ obtained in subsection 
III A.  This correspondence allows us to relate 
the first principles total energies and the derivatives of
total energies to various terms in $H_{eff}$.
The parameters in the quadratic part of $H_{eff}$ are linearly 
related to the force constant
matrices obtained from density functional linear response at high 
symmetry k-points in the BZ.
In Table (IX) are given specific relations
for modes at various k-points in the BZ including $\Gamma_{15}, 
X_2^{\prime},
X_5^{\prime}, M_5^{\prime}, M_2^{\prime}, R_{15}$ and the 
$\Lambda_{3}$ modes at $(111)\pi/(2a), (111)\pi/(4a)$. 
The parameter $\overline Z^*$ is determined from the eigenvector of 
unstable
$\Gamma_{15}$ and the Born effective charges.
Obtaining $\epsilon_\infty$ directly from DFT linear response
and solving the system of linear equations yields values for all  
the parameters in the quadratic part of $H_{eff}$, listed in
Table (IX). 
The resulting normal mode dispersion of $H_{eff}$ is shown in Fig. 3. 
For the LO modes at $(111)\pi/(2a)$ and $(111)\pi/(4a)$, the
 reasonable agreement between the calculated 
force constant matrix eigenvalue and $H_{eff}$ is an indication of 
the validity of the truncation in the Taylor expansion.

The parameters $B, C, D, E$ appearing in the onsite  anharmonic terms 
are determined from the total energies of uniformly distorted 
configurations ($\vec \xi_{i}=\vec \xi$), as shown in Fig. 4.
The minimum energy configuration has 
rhombohedral symmetry ($\vec \xi$ along (111)).
The difference among the energies of uniform distortions with 
different symmetries ((100), (110), (111)) is a 
reflection of the cubic anisotropy, which is described quite well by
the fourth order terms. The resulting parameters are listed in Table 
(X).
 
To account for the effects of changes in lattice parameters at the
structural phase transition,
we include the lowest order terms in the homogeneous strain and its 
coupling to the Pb-centered subspace:
\begin{eqnarray}
\nonumber
{N\over 2}C_{11}\sum_\alpha e_{\alpha \alpha}^2
+{N\over 2}C_{12}\sum_{\alpha \neq \beta}e_{\alpha \alpha}e_{\beta 
\beta}
+{N\over 4}C_{44}\sum_{\alpha \neq \beta}e_{\alpha \beta}^2
+Nf\sum_\alpha e_{\alpha \alpha}  \\
+g_0(\sum_\alpha e_{\alpha \alpha})\sum_i|\vec \xi_i |^2
+g_1\sum_\alpha(e_{\alpha \alpha}\sum_i\xi_{i \alpha}^2)
+g_2\sum_{\alpha < \beta}e_{\alpha \beta}\sum_i\xi_{i \alpha}\xi_{i 
\beta},
\end{eqnarray}
where $e_{\alpha \beta}$ is a component of the strain tensor, $C_{11},
 C_{12},
C_{44}$ are the elastic constants, and the parameters $g_0, g_1, g_2$
give the strength of coupling of strain with the local polar 
distortions
$\xi_{i \alpha}$. All these parameters are determined from the single
unit cell total-energy calculations for three independent types of
unit cell distortions (isotropic, uniaxial and rhombohedral shear),
with magnitudes of up to 2 to 4 $\%$ of
the experimental lattice constants. The total energies of these 
strained
unit cells with no local polar distortion, shown in Fig. 5,
give the three elastic constants $C_{11}, C_{12}$ and $C_{44}$.
For each of
these unit-cell-strain types, we also compute the second derivative 
of energy with respect
to uniform local polar distortions $\vec \xi_{i}=\vec \xi$, as shown 
in Fig. 5. 
These results yield the coupling parameters shown in Table (X).

Now we turn to the determination of the terms in $H_{eff}$ acting in 
the
Ti-centered subspace. Because this subspace contains the zone center 
acoustic
modes, these terms must satisfy global translational and rotational 
invariance.
This symmetry is built into the systematic expansion procedure 
given by Keating\cite{keat}, in which invariant terms are built up 
from dot
products of differences of the $\vec u_i$'s.
If the expansion of the effective Hamiltonian in the Keating 
construction\cite{keat} is truncated at quadratic order and 
three independent parameters, corresponding to
the three elastic constants, the following terms are obtained:
$${1\over N}\sum_i \tilde A|\vec u_i|^2 $$
\begin{eqnarray}
\nonumber
+{1\over N}\sum_i\sum_{\hat d=nn1}[ 
 \tilde a_L(\vec u_i\cdot \hat d)(\vec u_i(\hat d)\cdot \hat d)
+ \tilde a_T(\vec u_i\cdot\vec u_i(\hat d)
-(\vec u_i\cdot \hat d)(\vec u_i(\hat d)\cdot \hat d))]\\
{1\over N}\sum_i\sum_{\hat d=nn2}[
 \tilde b_L(\vec u_i\cdot \hat d)(\vec u_i(\hat d)\cdot \hat d)
+ \tilde b_{T1}(\vec u_i\cdot \hat d_1)(\vec u_i(\hat d)\cdot 
\hat d_1)
+ \tilde b_{T2}(\vec u_i\cdot \hat d_2)(\vec u_i(\hat d)\cdot 
\hat d_2)]
\end{eqnarray}
The relations of these parameters to the elastic constants are made by
 using
the Keating expansion to evaluate the energy of homogeneously strained 
configurations.
With these relations,
$\tilde A$ = $C_{11}$+$2C_{44}$, $\tilde a_L$ = $-{1\over 2}C_{11}$, 
$\tilde
a_T$ = $-{1\over 2}C_{44}$, $\tilde b_L$ = $-\tilde b_{T1}$ = 
$-{1\over 8}C_{12}$+${1\over 24}C_{44}$ and $\tilde b_{T2}$ = 0, these 
parameters can easily be
obtained from first-principles calculations. 
Because there are no unstable
modes in this subspace, there is no need to include higher-order 
interactions.
In any case, within the local anharmonicity approximation, global 
translational
invariance requires anharmonic terms to be zero at all orders.
As mentioned in the previous section,
there is no need for an explicit form of the 
Ti-centered LWF in this minimal treatment. 
For refinement of $H_{eff}$ in this subspace, one could construct an 
explicit form
and determine additional parameters in a manner analogous to that for 
the Pb-centered subspace.

Finally, the simplest coupling between local polar distortions 
(Pb-centered subspace)
and inhomogeneous strain (Ti-centered subspace) that satisfies the 
constraint of global
translational invariance and does not vanish in the limit $\vec k 
\rightarrow
0$ is the nearest-neighbor coupling linear in $\vec u$ and quadratic 
in $\vec \xi$, with both $\vec \xi$ variables taken on the same site:
\begin{eqnarray}
\nonumber
{\tilde h_0\over N}\sum_i\{ \xi_{ix}^2\sum_{\vec d = \pm\hat
y\pm\hat z}(u_x(\vec R_i + {a_0\over 2}\hat x + {a_0\over 2}\vec d)-
u_x(\vec R_i - {a_0\over 2}\hat x + {a_0\over 2}\vec d))+c.p.\}\\ 
\nonumber
+{\tilde h_1\over N} \sum_i\{\xi_{ix}^2(\sum_{\vec d = \pm\hat x\pm
\hat z}(u_y(\vec R_i + {a_0\over 2}\hat y + {a_0\over 2}\vec d)-u_y
(\vec R_i -
{a_0\over 2}\hat y + {a_0\over 2}\vec d))\\ 
\nonumber
+\sum_{\vec d = \pm\hat x\pm\hat y}(u_z(\vec R_i + {a_0\over 2}\hat z 
+ {a_0\over 2}\vec d)-u_z(\vec R_i - {a_0\over 2}\hat z + {a_0\over 2}
\vec d)))+c.p.\}\\ 
\nonumber
+{\tilde h_2 \over N}\sum_i\{\xi_{ix}\xi_{iy}(\sum_{\vec d = \pm\hat x
\pm\hat
z}(u_x(\vec R_i + {a_0\over 2}\hat y + {a_0\over 2}\vec d)-u_x(\vec R_i
 - {a_0\over 2}\hat y + {a_0\over 2}\vec d))\\ 
+\sum_{\vec d = \pm\hat y\pm\hat z}(u_y(\vec R_i + {a_0\over 2}\hat x 
+ {a_0\over 2}\vec d)-u_y(\vec R_i - {a_0\over 2}\hat x + {a_0\over 2}
\vec d)))+c.p.\}
\end{eqnarray}
$\xi_{i\alpha}^2$ couples only to differences of the $u_{\beta}$'s, 
which can be recognized as
finite difference approximations to the gradient, and thus as the local 
strain tensor (see also Ref. \onlinecite{zvr}).
The terms in $H_{eff}$ coupling
inhomogeneous strain and polar distortions are related in the long 
wavelength
limit to the coupling between homogeneous strain and the polar 
$\Gamma_{15}$ 
mode. Thus the three independent coupling parameters can be obtained 
from the corresponding homogeneous strain coupling parameters:
$\tilde h_0 = (g_0+g_1)/4, \tilde h_1 = g_0/4$ and $\tilde h_2=g_2/8$.

\subsection{Examination of model energetics}

Having fully determined $H_{eff}$,
we now explore the low energy surface of the model to confirm 
that it gives a correct ground state when compared with the
real crystal. Since the anharmonic terms occur only in
the Pb-subspace and are local (the anharmonicity is wavevector  
independent), it is easy to determine the ground state from the 
quadratic order terms.
The lowest energy mode is obtained by freezing in the most 
unstable mode: $\Gamma_{15}$. We consider changes in energy as this 
mode is frozen in with polarization along the $(001), (110)$ and 
$(111)$ directions. In Fig. 4, it can be seen that the rhombohedral
state ($(111)$-distortion) has the lowest energy. If the unit cell is 
allowed
to relax as the mode is frozen in, by minimizing over the homogeneous
 strain,
we find an overall increase in distortion energy, with the lowest 
energy state 
being of tetragonal symmetry ($(001)$-distortion) as can be seen in 
Fig. 6. This is consistent both
with previous first principles calculations\cite{ferroc} and 
experimental results\cite{shirane1}.
 
For the lowest energy tetragonal configuration, we obtain
a value for the spontaneous polarization from the mode effective 
charge of $P_s = 0.87 C/m^2$, which is in the range of
values $(P_s = 50$ to $100 C/m^2)$ reported from experiments\cite{lg}. 
From the values of homogeneous strain in this ground state, we obtain 
a c/a ratio of 1.08, to be compared with the experimental value of 
1.061\cite{shirane1}.
Using the explicit form for the LWF, the atomic displacements in the 
model ground state can be obtained.
We find that the oxygen
octahedra are almost undistorted and the relative displacement of the 
Pb atoms is 0.35 \AA (to be compared with the experimental value of 
0.47 \AA\cite{shirane1}).
It should be noted that this is not the true LDA ground state, since 
previous total energy calculations\cite{rw2} for the experimental
distortions showed the latter is slightly lower in energy.
This means that 
there is a higher order coupling of the unstable $\Gamma_{15}$ mode to
an additional polar $\Gamma_{15}$ mode not included in the effective 
Hamiltonian subspace. Considering that the atomic displacements in 
$PbTiO_3$ 
are relatively large, the presence of such anharmonicity is not 
surprising.
However, as discussed above, the ground state of the model is very 
similar to
the experimental ground state and the small loss of accuracy is more 
than
compensated for by the gain in simplicity.

\section{Finite-temperature behaviour}

The effective Hamiltonian is constructed to show the same
finite temperature critical behaviour as the full lattice Hamiltonian
in a statistical mechanical analysis.
While the form of $H_{eff}$ is somewhat too complicated for the 
application
of analytical methods beyond mean field theory, it is quite suitable 
for
Monte Carlo simulations, since the changes in energy for changing
 system configurations can be readily evaluated. 
Monte Carlo simulations are used in the detailed analysis of $H_{eff}$ 
to obtain T-dependence of a variety of structural properties near the 
transition,
while our mean field theory analysis is limited to the estimation of 
$T_c$ and
the identification of the order of the transition and symmetry of the
 phases.
Comparison of the mean field results with those of Monte Carlo 
simulations allows us to study the effects of fluctuations.

\subsection{Mean field theory}

Variational mean field theory for the class of models with variable 
length vector
degrees of freedom and strain coupling was developed in Ref. 
\onlinecite{royal}.  In this approach,
the homogeneous strain and uniform polarization ($\vec P$) are 
identified as the order parameters for the transition. 
$\vec P$ is directly related to the average value
of uniform local distortion through the mode effective charge 
$\overline Z^{\star}$ and the unit cell volume $\Omega_{cell}$,

\begin{equation}
\vec P = \overline Z^{\star}.<\vec\xi>/\Omega_{cell}.
\end{equation}

In the high temperature (cubic perovskite) phase, the uniform 
polarization is zero 
and the strain tensor has full cubic symmetry: 
$e_{xx}=e_{yy}= e_{zz}$. We used the variational formulation of mean 
field theory, which involves
constructing a trial density matrix as a product of single site 
density matrices and 
minimizing the resulting free energy functional with respect to the 
variational parameters in the trial density matrix.
We minimized the trial free energy with respect to variational 
parameters corresponding to cubic, tetragonal and rhombohedral 
symmetries to determine the
stable phase at various temperatures. The system is stable 
in the cubic phase above the transition temperature $T_c = 1100 K$ 
and 
in the tetragonal phase below $T_c$, but within the accuracy of our
calculation, the transition is second order. Switching off the 
coupling
to homogeneous strain resulted in a second order cubic-rhombohedral 
transition at a significantly lower temperature of 900 K.

\subsection{Monte Carlo simulations}

Classical Monte Carlo simulations\cite{alt} were performed using the 
Metropolis
algorithm for finite size systems of $L \times L \times L$ unit cells 
and periodic boundary conditions.
A configuration
of the system is specified by two sets of three dimensional
vectors $\{\vec \xi_{i}\}$ and $\{\vec u_{i}\}$ placed on 
interpenetrating 
simple cubic lattices of size $L \times L \times L$. 
We generated a trial configuration by updating a single vector 
to a randomly chosen vector inside a cubic box 
centered at the current value of the vector. The size of this box is 
chosen
to yield a reasonable acceptance ratio $(>0.1)$, and is roughly 
$0.25 a_0$
near $T_c$.  With the change in a single vector, the change in energy 
associated with short range terms (quadratic interactions up to 
third neighbour, the onsite potential, coupling to strain and 
third order coupling between the two subspaces) is easy to calculate.
Because of their long range, computation of dipolar intersite 
interactions is 
relatively costly, limiting the size of our simulations to $L\leq 
12$.
The $3 \times 3$ matrix of dipolar intersite coupling for each pair 
of
 spins
was calculated using the Ewald summation technique for each value of 
L.
Changes in the quadratic intersite interaction due to changes in 
strain
are neglected in this model.
One Monte Carlo sweep (mcs) involves one 
update of the $\xi_i$'s (in typewriter mode) followed by 
one update of the $u_i$'s (in typewriter mode), and 20 updates of 
the 6 components of the strain tensor.

Preliminary Monte Carlo simulations performed with 25,000 to 50,000 
sweeps 
showed dependence on the initial configuration of simulations at 
temperatures in the vicinity of the transition for $L > 5$. 
For $L = 5$, which is small enough for ergodic sampling in a run of 
this
length, the energy histogram shows two clearly separated peaks.
This behaviour is typical of a first order transition\cite{multc}.
At larger system sizes, 
due to the exponentially increasing free energy barrier between the 
regions of 
configuration space corresponding to low and high temperature phases, 
only one of the two peaks in the energy histogram is sampled, 
depending on the choice of initial configuration.
An accurate determination of $T_c$ requires knowledge of the relative
free
energies of the high and low temperature phases as a function of 
temperature.
Recently developed methods to extract $T_c$ for first order 
transitions 
include multicanonical algorithms\cite{multc} and jump-walking 
algorithms\cite{jump}.  In our applications of these methods to 
$PbTiO_3$, 
we found that these approaches require very long ($10^6$ mcs) 
simulations, 
and therefore are rather impractical.
However in the present case, the uncertainty in $T_c$ obtained from
 the 
range of hysteresis is small compared to the LDA and other errors in 
our 
analysis and therefore high accuracy determination of $T_c$ is not 
necessary.
The calculation of the physical properties of the high and low 
temperature phases at temperatures inside the range of hysteresis is 
carried out with an appropriate choice of the initial configuration.

In Fig. 7, we show 
the bounds on $T_c$ for $L=5$ through $11$, obtained by monitoring the 
sensitivity of the average
structural parameters to the choice of initial state: $T_>$ is the 
lowest temperature at which the
system averages are characteristic of the cubic state, starting with
 an inital
ground state tetragonal configuration, while $T_<$ is the highest 
temperature
at which a starting cubic configuration results in a tetragonal state. 
A value of $T_c=660 K$, obtained from averaging the bounds at the 
largest
system size, is in very good agreement with the experimental 
transition
temperature 763K. 

To detect the symmetry of the low temperature phase, we calculated the
 averages
of largest, smallest and intermediate absolute values of the cartesian 
components of $\vec\xi=\frac{1}{N} \sum_i(\vec\xi_i)$. These averages
 for the $7 \times 7 \times 7$ system as a function of temperature 
are shown in Fig. 8. Near $T_c$,
the largest component jumps to a finite value, while the other two 
components remain close
to zero, indicating tetragonal symmetry of the low temperature phase. 
As shown in Fig. 9, this uniform tetragonal polar distortion is 
accompanied
by a tetragonal strain $c/a \neq 1$. This quantity also shows a marked
 jump near $T_c$.
Finally, from the average homogeneous strain we obtained the average
volume of the system as a function of temperature, as shown in 
Fig. 10.
The negative thermal expansion in the simulations just below $T_c$
is also seen experimentally\cite{volume}.

The latent heat of a first order transition is given by the difference
in energies at which the two peaks appear in the energy 
histogram in the simulations at $T_c$. 
To determine this energy difference,
we performed two simulations for $L=9$ at the midpoint of the
 hysteresis range 
$T_c=675 K$, one starting with a tetragonal configuration
and the other starting with a cubic configuration. The difference
in the positions of the peaks in the energy histograms  for these two
simulations yields an estimate of latent heat of 3400 J/mol, in rough 
agreement with the measured value of 4800 J/mole\cite{shirane2}. 
It should be noted that these values are much larger than 209 J/mol 
latent heat of the cubic tetragonal transition of 
$BaTiO_3$\cite{shirane3}. 

Information about the local distortions in the high temperature
nonpolar phase just above $T_c$, can be obtained from
the single spin distribution $<\vec\xi_i>$. 
For all $L$, we find the distribution to be very close to Gaussian.
The rather broad width ($\approx 0.04 a_0$) shows that there are 
significant local distortions.

For the system sizes used in the simulations, the coupling to the 
inhomogeneous
strain appears to be relatively unimportant. If the coupling is set to 
zero
the changes in the calculated $T_c$ and other transition properties 
are
negligible. However, for larger scale simulations involving multiple
domains the effects should be significant.

\section{Discussion}

The first principles effective Hamiltonian
constructed in the previous section provides a quantitative 
microscopic
description of the structural energetics of $PbTiO_3$ relevant to the 
paraelectric-ferroelectric phase transition. This model can be used to
investigate the connection of specific features of the Hamiltonian to 
the
observed behaviour in the vicinity of the transition. In addition, it 
is 
possible to connect these features to aspects of the chemistry of 
$PbTiO_3$
and related compounds.

One important feature of the Hamiltonian is that the TO branches are 
unstable
throughout most of the BZ (Fig. 3). So, although $\Gamma_{15}$ is the 
dominant
instability, finite wavelength fluctuations have relatively low 
energy. This
may account for the breadth of the single site distribution, and can 
be expected also to be reflected in the short range order. 
The unstable branch along the $(111)$ direction in $PbTiO_3$ is quite
flat when compared with $SrTiO_3$\cite{chris} and $KNbO_3$\cite{yuk}.
In comparison with $KNbO_3$ and ferroelectric $BaTiO_3$, in which 
the polar unstable modes have a strong B-component\cite{yuk,zvr}, the 
instabilities in $PbTiO_3$ are dominated by large Pb-displacements. 
From symmetry arguments\cite{rwwannier}, the Pb displacements couple 
with oxygen displacements
at $\Gamma, X, R$ and $M$ leading to the low energy of the modes at
 those points.
While the same argument applies to Ti displacements at $\Gamma, X$ 
and $M$, they do not couple with any other mode at $R$ point.
Therefore, the energy of the mode at $R$ is high and the dispersion 
along $\Gamma$ to R is large.  The special role $Pb$ plays in
the instabilities of $PbTiO_3$ and $PbZrO_3$\cite{wrpz} in contrast 
with the 
A-atoms in non-$Pb$ perovskite compounds has its origin in the strong 
hybridization of $Pb$ with oxygen atoms\cite{ferroc,dsingh}. 

To understand the consequences of the coupling of the polar subspace 
to the 
strain at finite temperature, we performed Monte Carlo simulations for
 $H_{eff}$
with this coupling set to zero. This corresponds to a constant volume
phase transition with the unit cell constrained to be cubic. In this 
case,
we find a second order phase transition at $400$ K directly to the
rhombohedral phase. Thus, the coupling of local polar distortions to 
strain is responsible for both the stability of the tetragonal 
phase relative to the rhombohedral one and the first order character
 of the 
transition at finite temperature. As discussed in the previous 
section,
mean field theory shows a second order transition both with and 
without 
strain coupling, implying that both fluctuations and strain coupling 
are
required for producing the first order transition. 
Comparing the transition temperatures obtained in Monte Carlo and mean
 field theory with and without strain coupling, we find that 
while fluctuations suppress $T_c$, the coupling to strain enhances the 
stability of the ferroelectric phase.

In comparison with related ferroelectric compounds, the transition 
in $PbTiO_3$
has a much stronger first order character, reflected in its large 
latent heat.
While the strain coupling is responsible for the first order 
character\cite{strainw},
anharmonicity in the lattice plays an important role in the magnitude 
of its
discontinuity\cite{royal}. The minimum energy uniform polar 
distortions in
$PbTiO_3$ are much larger than those in related compounds indicating 
a large 
contribution from anharmonicity in the low-energy surface. The 
relation of 
these
features to the chemistry of A or B atoms was discussed using 
ionic radii in Ref. \onlinecite{royal}.

There are two main sources of error in the work presented in this 
paper. One of these is the LDA used in the exchange correlation 
functional.
Equilibrium lattice constants are typically underestimated in the LDA
calculations. This can strongly affect the study of structural phase 
transition,
since the lattice instabilities are very sensitive to the lattice 
parameters.
In the present work these errors were partially eliminated by 
expanding the effective Hamiltonian around 
the experimental lattice constant near $T_c$ and dropping the term 
linear in strain.
The other source of error is the truncation of the effective 
Hamiltonian
subspace. In the LWF approach, this subspace is decoupled 
at quadratic order from its complementary subspace to a good 
approximation. However there can be anharmonic
coupling between the two subspaces. In the case of 
$PbTiO_3$, there is a small higher 
order coupling of $\Gamma_{15}$ to modes not included in the subspace,
which affects the energies of large distortions.
Since these
large distortions are mainly important at low temperatures, we expect 
this coupling to have a small effect on $T_c$. 

\section{Conclusion}
In conclusion, we have applied the method of lattice 
Wannier functions to construct an effective Hamiltonian for the 
ferroelectric
phase transition in $PbTiO_3$ completely from first principles. 
Monte Carlo simulations for this Hamiltonian yield a first order
cubic-tetragonal transition at 660 K and a description of the system 
near the
transition in good agreement with experiment. The strong involvement 
of Pb atom
in the lattice instability as well as anharmonicity and the coupling 
of polar
distortions with homogeneous deformations of the lattice are found to
 be very
important in producing the transition behaviour characteristic of 
$PbTiO_3$.

\acknowledgements

We are grateful for useful discussions with R. E. Cohen, E. Cockayne,
B. A. Elliott, Ph. Ghosez and Serdar Ogut.
We thank M. C. Payne and V. Milman for the use of and valuable 
assistance
with CASTEP 2.1. This work was supported by ONR Grant N00014-91-J-
1247.  Part of the calculations were performed at the Cornell 
Theory Center.
In addition, K. M. R. acknowledges the support of the Clare Boothe 
Luce Fund and the Alfred P. Sloan Foundation.

\begin{table}
\caption{Cubic perovskite lattice and elastic constants calculated
from various first principles calculations. Elastic constants are 
given in eV/cell.}
\begin{tabular}{lccc}
 & This work & Ref. \protect\onlinecite{ferroc}  & 
Ref. \protect\onlinecite{ksv} \\
\tableline
      $a_0$ (\AA)&    3.883 &  3.889 & 3.889    \\
      B (GPa)&   203    & 215    & 209      \\
     $C_{11}$   &  117. &  -     & 123.  \\
     $C_{12}$   &  51.6 &  -     & 53.6  \\
     $C_{44}$   &  137. &  -     & 148.  \\
\end{tabular}
\end{table}

\begin{table}
\caption{Effective charges calculated from first
principles linear response and compared with the
results of the geometric phase approach (Ref. 
\protect\onlinecite{zkv}).}
\begin{tabular}{lcccc}
 &$Z_{pb}^{\star}$ & $Z_{ti}^{\star}$ & $Z_{o1}^{\star}$ & 
$Z_{o2}^{\star}$ \\
\tableline
This work & 3.87 & 7.07 & -5.71 & -2.51  \\
Ref. \protect\onlinecite{zkv} & 3.90 & 7.06 & -5.83 & -2.56 \\
\end{tabular}
\end{table}

\begin{table}
\caption{IR active optical phonon frequencies ($cm^{-1}$)
at $\Gamma$ obtained using linear response at the experimental volume.
 They are compared with the 
results of the frozen phonon calculations performed at the LDA volume
 with ultrasoft pseudopotentials \protect\cite{zkv}.}
\begin{tabular}{lcccccc}
     & TO1   &  TO2  &   TO3  &  LO1  &   LO2  &   LO3  \\
\tableline
 Present work & 182 I  &  63   &  447   &  47   &   418  &   610 \\

 Ref. \protect\onlinecite{zkv}  & 144 I  & 121   &  497 & 104 & 410 
& 673  \\
\end{tabular} 
\end{table}

\begin{table} 
\caption{LO-TO splitting: mode effective charges and correlation 
matrix.}
\begin{tabular}{lcccc}
      &  $\overline Z^{\star}$ &    LO1    &   LO2    &  LO3  \\  
\tableline
 TO1  &  9.45 & 0.224   &  0.466   & 0.855 \\
 TO2  &  2.56 &   0.974 &    0.116 &   0.192 \\
 TO3  &  1.53 &  0.010  &   0.876  &  0.481 \\
\end{tabular} 
\end{table}

\begin{table}  
\caption{Selected phonon frequencies ($cm^{-1}$) at high symmetry 
k-points calculated using DFT linear response. Symmetry labels follow 
the convention of Ref. \protect\onlinecite{rwwannier}. }
\begin{tabular}{lcc}
k-point &  phonon  &  frequencies \\
\tableline
X  &   $X_5^{\prime} $ & 30.6 I, 264 \\
   &   $X_2^{\prime} $ & 93.1, 647   \\
\tableline
M & $M_5^{\prime}$ & 35.1 I, 400, 201 \\
  & $M_2^{\prime}$ & 16.4 \\
\tableline
 & $R_{25}$  & 145 I   \\
 & $R_{15}$  & 15.5, 339 \\
R & $R_{25}^{\prime}$ & 367 \\
 & $R_{12}^{\prime}$ & 370 \\
 & $R_2^{\prime}$ & 746 \\
\tableline
   & $\Lambda_1$  &  8.78, 249, 421, 696 \\
(111)$\frac{\pi}{2a}$   & $\Lambda_2$  &  148 \\
   & $\Lambda_3$  &  58.2 I, 82.9, 230, 301, 430 \\
\end{tabular} 
\end{table}

\begin{table}  
\caption{Determination of LWF parameters. Linear combinations of these 
parameters for the modes at high symmetry k-points and the 
corresponding components of the
normalized eigenvectors of the force constant matrix.}
\begin{tabular}{lcc}
Mode &  combination of the parameters & component of the eigenvector\\
\tableline
$\Gamma_{15}$  &  $p_1+4 p_2+2 p_3+12 p_4$ & 0.5560  \\
               & $8 t_1$ & 0.5375        \\
               & $4 o_1$ & -0.3414         \\
               & $4 o_2$ & -0.4109     \\
\tableline
$R_{15}$       & $p_1-4 p_2-2 p_3+12 p_4$ & 0.8981 \\
               & $4 o_3-8 o_6$ & -0.3110        \\
\tableline 
$M_{2}^{\prime}$ & $p_1-4 p_2+2 p_3-4 p_4$ & 1.0000  \\
\tableline 
$M_5^{\prime}$ & $p_1-2 p_3-4 p_4$ & 0.9010  \\
               & $8 t_2$ & 0.3024 \\
\end{tabular}
\end{table}

\begin{table}  
\caption{Values of the LWF parameters determined from first 
principles. 
The parameters of the approximate LWF described in the text are given
 in the second column. 
Parameters for the refined LWF are obtained
by fitting to all the TO modes at $\Gamma, R, X$ and $M$, with 
additional parameters
associated with third neighbour shell of Pb atoms and second neighbour
 shell of oxygen atoms.}
\begin{tabular}{lcc}
Parameter &  Approx. LWF   & Refined LWF  \\
\tableline
$p_1$ &     0.839     &      0.829  \\
$p_2$ &    -0.037    &     -0.049  \\
$p_3$ &    -0.012    &      0.014  \\
$p_4$ &    -0.009   &     -0.019  \\
$p_5$ &     0.0          &      0.017  \\
$o_1$ &    -0.085    &     -0.086  \\
$o_2$ &    -0.102     &     -0.103  \\
$o_3$ &    -0.077    &     -0.087  \\
$o_4$ &   0.             &          .00045 \\
$o_5$ &   0.             &         -.0045  \\
$t_1$ &     0.067     &      0.067  \\
$t_2$ &     0.038    &      0.037  \\
\end{tabular}               
\end{table}

\begin{table}  
\caption{Comparison of $X_5^{\prime}$ eigenvectors. Mode vector 
(first row) built up using 
the approximate LWF is compared with the eigenvector (second row) of 
the force constant matrix at X.}
\begin{tabular}{lcc}
    &   Pb component   &   O component   \\
\tableline
Mode in the subspace    &  0.853 &  -0.341  \\
Eigenvector from LR              &  0.937 &  -0.349  \\
\end{tabular}               
\end{table} 

\begin{table} 
\caption{Determination of coefficients in the quadratic part of 
$H_{eff}$.  Linear combinations of these coefficients for the modes 
in the $H_{eff}$
subspace at high symmetry k-points are equated to the corresponding
eigenvalues of the projected force constant matrix.}
\begin{tabular}{lcc}
k-point     &  Mode eigenvalue at k of the effective Hamiltonian  &  
 Value from LR \\
           &                   &        ($eV/\AA^2$) \\
\tableline
 -    &       $z=\overline Z^{\star2}/\epsilon_{\infty}$          &                         12.18     \\

$\Gamma_{15}$ & $ A+2 (a_L+2 a_T)+4 (b_L+b_{T1}+b_{T2})+8 (c_L+2 c_T)
/3 -0.964843 z/2$ & -1.908 \\

$X_{2}^{\prime}$ & $ A-2 a_L+4 a_T-4 (b_L+b_{T1})+4 b_{T2}-8 (c_L+2 
c_T)/3+ 2.231399 z/2$ & 6.467 \\
 
$X_{5}^{\prime}$  & $A+2 a_L-4 b_{T2}-8 (c_L+2 c_T)/3-1.115699 z/2$ &       
           -0.266  \\

$M_{5}^{\prime}$  & $ A-2 a_L-4 b_{T2}+8 (c_L+2 c_T)/3+0.6165696 z/2$ 
&   -0.360   \\

$M_{2}^{\prime}$ &  $A+2 a_L-4 a_T-4 (b_L+b_{T1})+4 b_{T2}+8 (c_L+2 
c_T)/3 -1.23314 z/2$ &  0.103  \\

$R_{15}$ & $ A-2 a_L-4 a_T+4 (b_L+b_{T1})+4 b_{T2}-8 (c_L+2 c_T)/3$ &  
             0.076     \\

$(111)\frac{\pi}{2a}$  & $A-(-2 (b_L-b_{T1})+0.41635523 z/2.0)$ &   
                          -0.568  \\

$(111)\frac{\pi}{4a}$  & $ A+\sqrt{2} (a_L+2 a_T)+2 (b_L+b_{T1}+
b_{T2}) +0.942809 (c_L+2 c_T)$&   \\
          & $-((-b_L+b_{T1})-0.942809 (c_L-c_T)+0.7953677 z/2)$  &
                   -1.750    \\
\end{tabular}               
\end{table}

\begin{table}
\caption{Parameters in the effective Hamiltonian (units of eV per 
unit cell).}
\begin{tabular}{||c|r||c|r||c|r||}
\hline $~~A~~$     &  18.43~  &$a_L\qquad$  &  39.27~  &  
$C_{11}\qquad$   &
117.9~\\
\hline$B$     &  2.629$\times10^3~$  &   $a_T\qquad$   & -10.67~  &
$C_{12}\qquad$   &  51.50~ \\
\hline$C$     &  4.277$\times10^3~$  &   $b_L\qquad$   &   4.882~  &
$C_{44}\qquad$   &  137.2~
\\  \hline$D$     &  -1.658$\times10^5~$  &   $b_{T1}\qquad$&  
-1.391~ &
$g_{0}\qquad$    & -107.7~
\\  \hline$E$     &  9.630$\times10^6~$  &   $b_{T2}\qquad$&  -0.1434~ 
 & $g_{1}\qquad$    & -790.3~  \\
\hline$~\overline Z^{*2}/\epsilon_\infty~$     & 12.18~  &   
$c_{L}\qquad$ & -3.389~  &  $g_{2}\qquad$    & -357.09~  \\
\hline        &           &   $c_{T}\qquad$ &   0.7104~  &  $f\qquad$
        & 4.48~  \\
\hline\hline
\end{tabular}
\end{table}

\begin{figure}
\caption{(a) Unit cell of the cubic perovskite compounds $ABO_3$
(b) Low temperature crystal structure of $PbTiO_3$. Displacements 
of the atoms indicated by arrows form the polar distortions of the
cubic unit cell.}
\end{figure}

\begin{figure}
\caption{$z$ component of the vector-like Pb-centered lattice Wannier 
functions. $Pb, Ti$ and $O$ atoms are represented by solid squares,
empty squares and circles respectively.
Parameters labeling the displacement patterns correspond to
the length of the displacements (arrows) of atoms for the unit value 
of the LWF coordinate.}
\end{figure}

\begin{figure}
\caption{Normal mode dispersion of $H_{eff}$. Solid circles
are the first principles mode eigenvalues used in the fitting.
Open circles are the first principles mode eigenvalues not used
in fitting the $H_{eff}$, which test the validity of the truncated 
form of the effective Hamiltonian.}
\end{figure}

\begin{figure}
\caption{Total energies for uniformly distorted configurations ($\vec 
\xi_i= \vec \xi$) along directions $(001), (110)$ and $(111)$. Solid 
lines are
the fit obtained with the parameters $B, C, D$ and $E$ in $H_{eff}$.}
\end{figure}

\begin{figure}
\caption{Energetics of the homogeneous strain ( (a) isotropic, (b) 
uniaxial
and (c) shear) and its coupling to the uniform polar distortions.
Circles are the total energies for the strained unit cell 
configurations with no
polar distortions. Solid lines going through the circles are the fits 
obtained with the elastic constants $C_{11}, C_{12}$ and $C_{44}$. 
Squares correspond to the second derivative of the total energies 
with respect to uniform polar distortions for the strained unit cells.
Solid lines going through the squares are the fits obtained with the 
coupling parameters $g_0, g_1$ and $g_2$.}
\end{figure}
 
\begin{figure}
\caption{Model energetics of the uniform polar distortions along 
$(100), (110)$ and $(111)$. Dotted lines correspond to the polar 
distortions with the unstrained cubic unit cell,
and solid lines to the distortions with unit cell allowed to relax 
with respect to homogeneous strain.}
\end{figure}

\begin{figure}
\caption{Bounds on the transition temperature $T_c$ as a function of 
system size used in the simulations.}
\end{figure}

\begin{figure}
\caption{Averages of largest, smallest and intermediate absolute 
values of the cartesian components of the order parameter $\vec \xi 
= \sum_i \vec \xi_i/N$ as a function of temperature, obtained from 
Monte Carlo simulations (125000 mcs) of a $7 \times 7 \times 7$ 
system.}
\end{figure}

\begin{figure}
\caption{Average tetragonal strain parameter $\frac{c}{a}$ as a 
function of 
temperature, obtained from Monte Carlo simulations (125000 mcs) of 
a $ 7 \times 7 \times 7$ system.} 
\end{figure} 

\begin{figure}
\caption{Average unit cell volume as a function of temperature, 
obtained from Monte Carlo simulations (125000 mcs) of a $7 \times 
7 \times 7$ system.}  
\end{figure}  

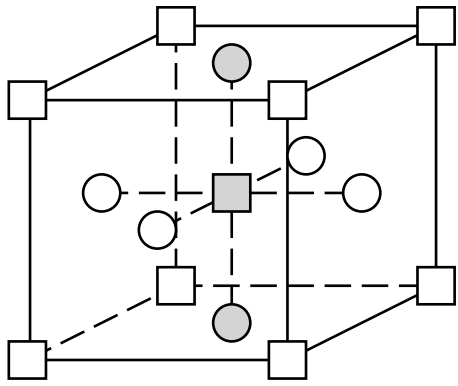
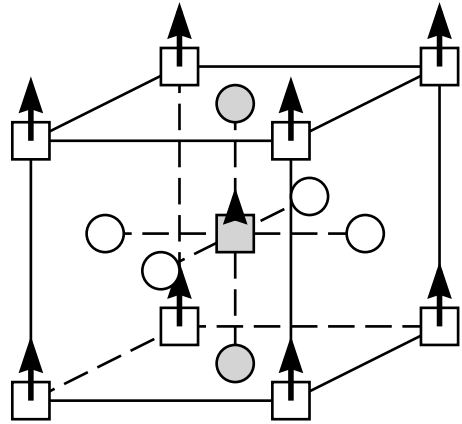
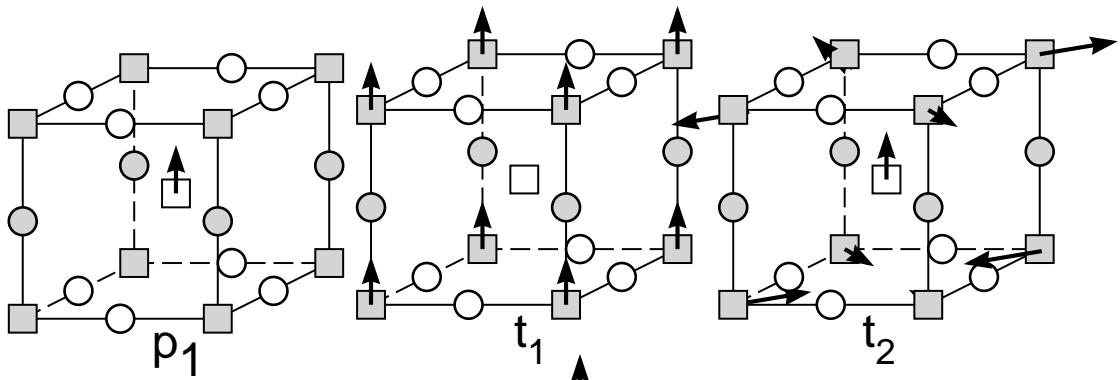
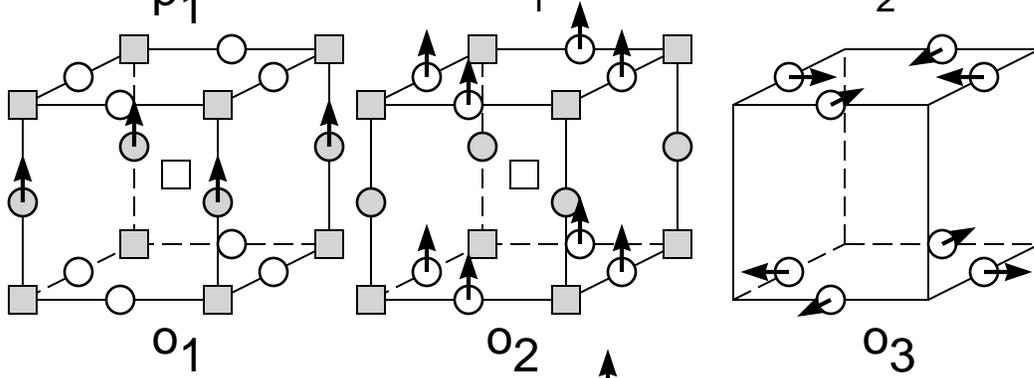
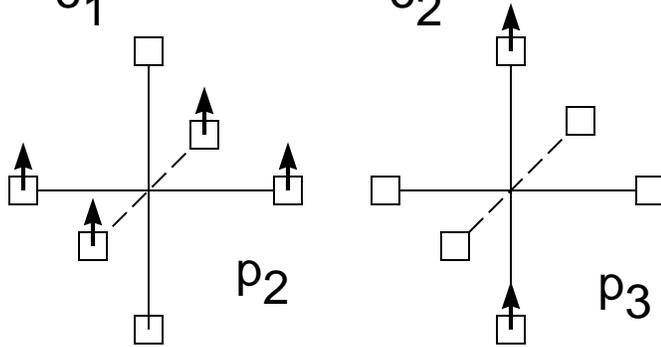
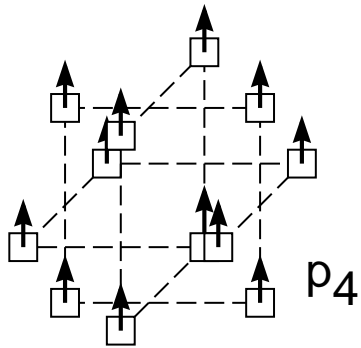
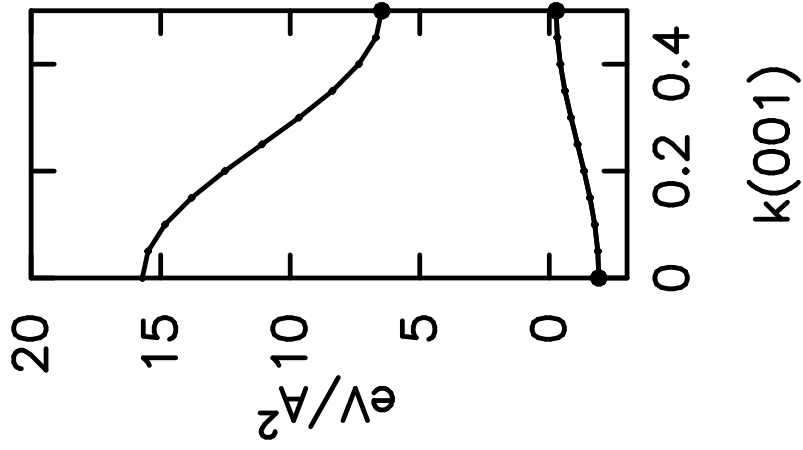
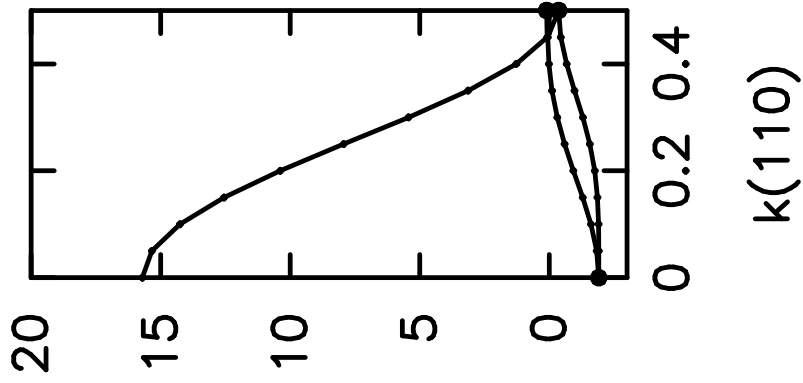
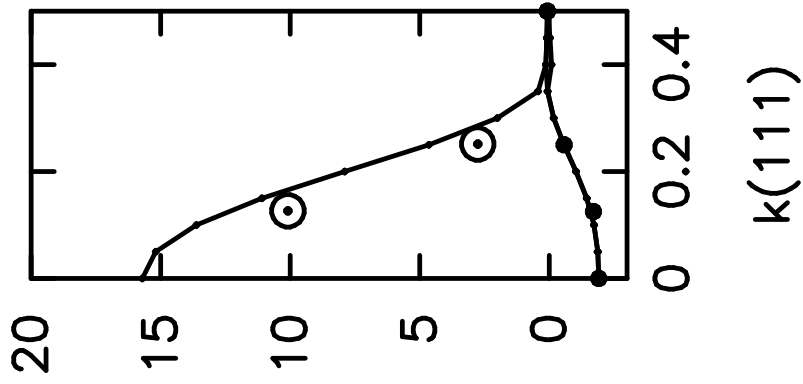
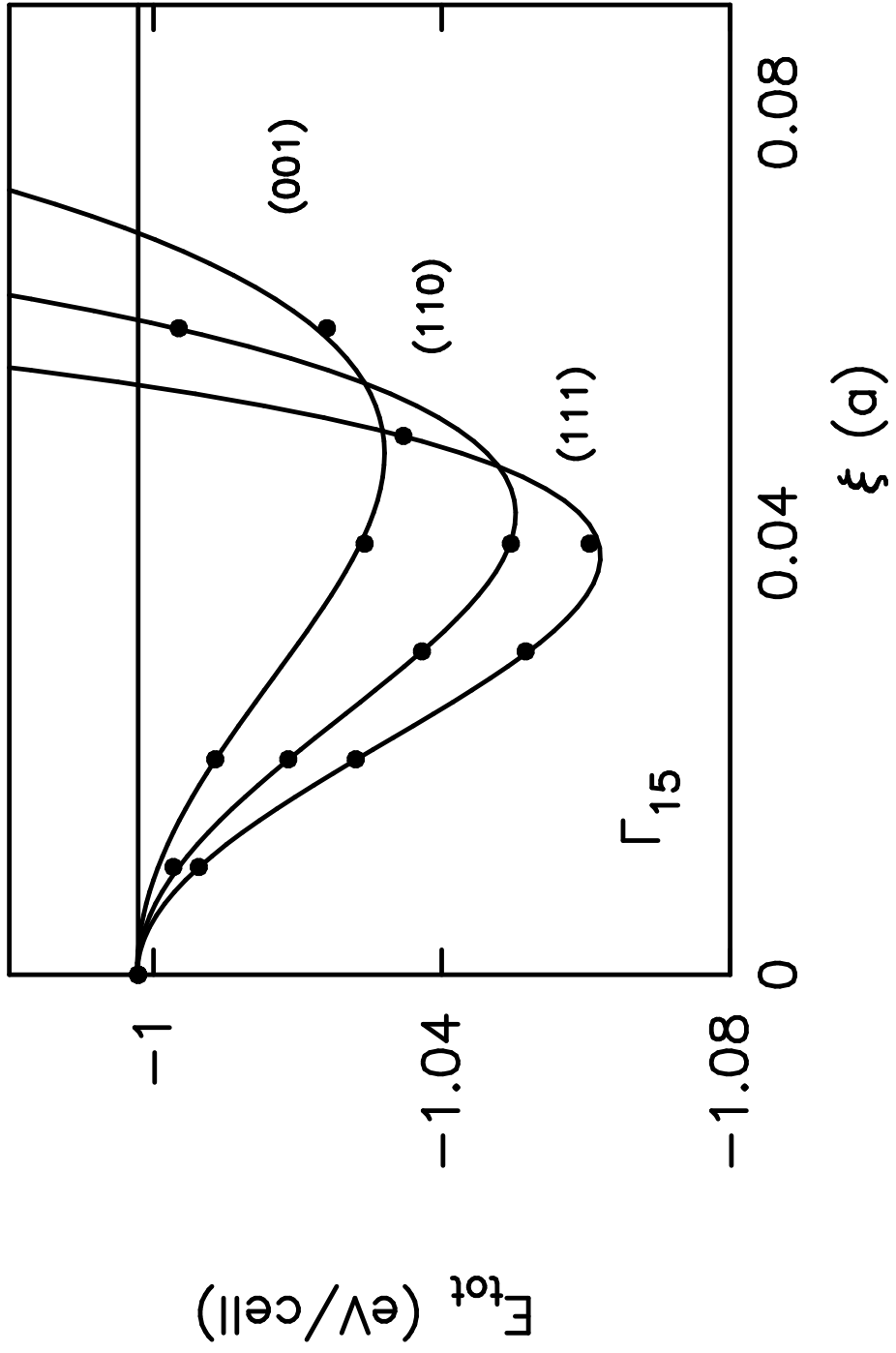
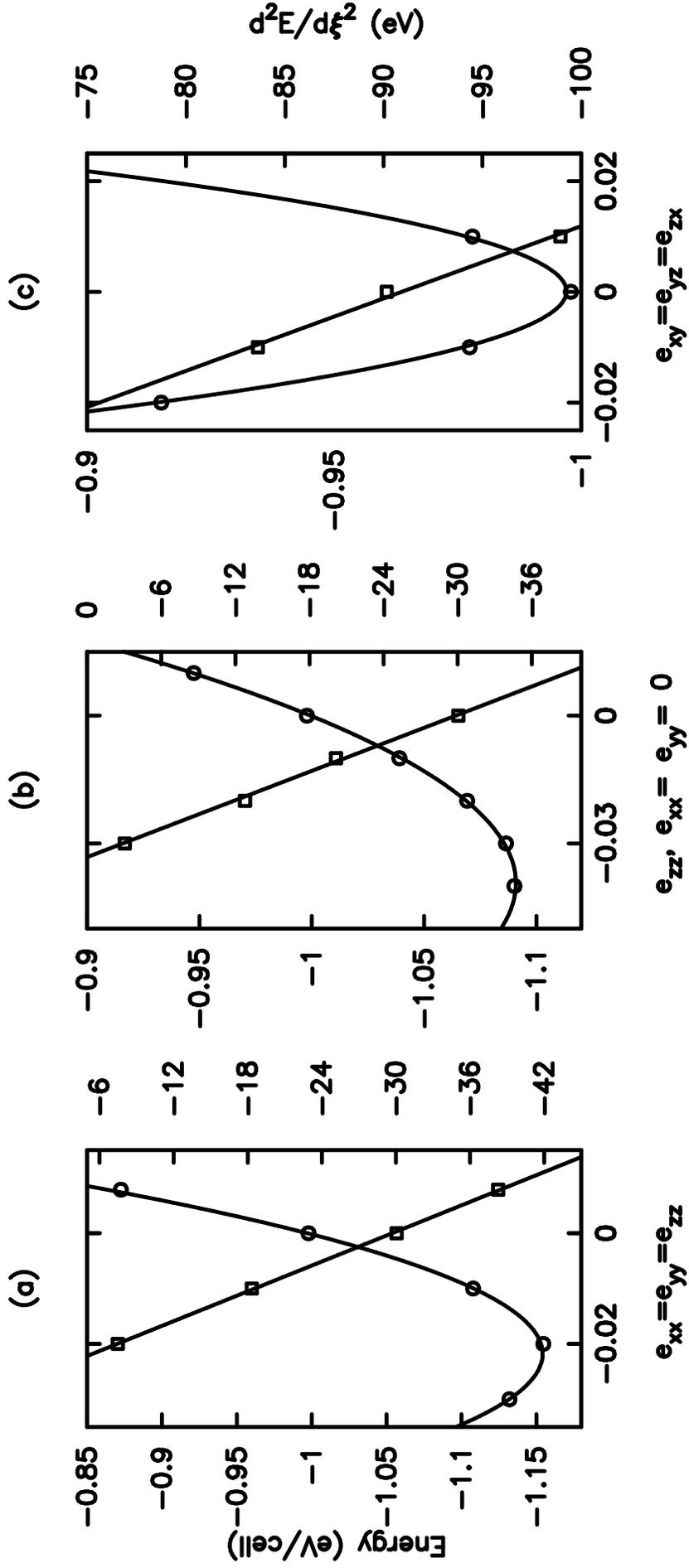
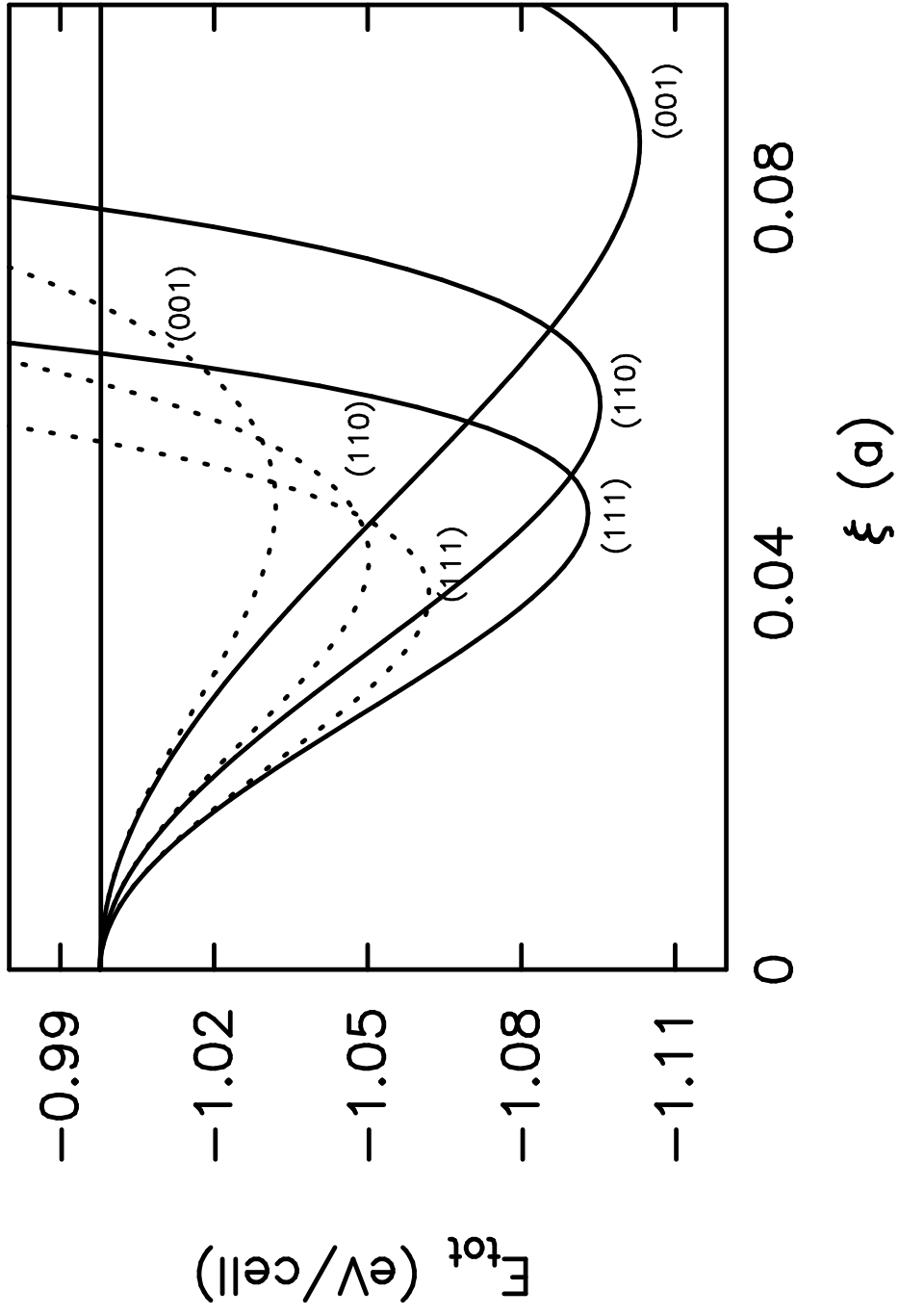
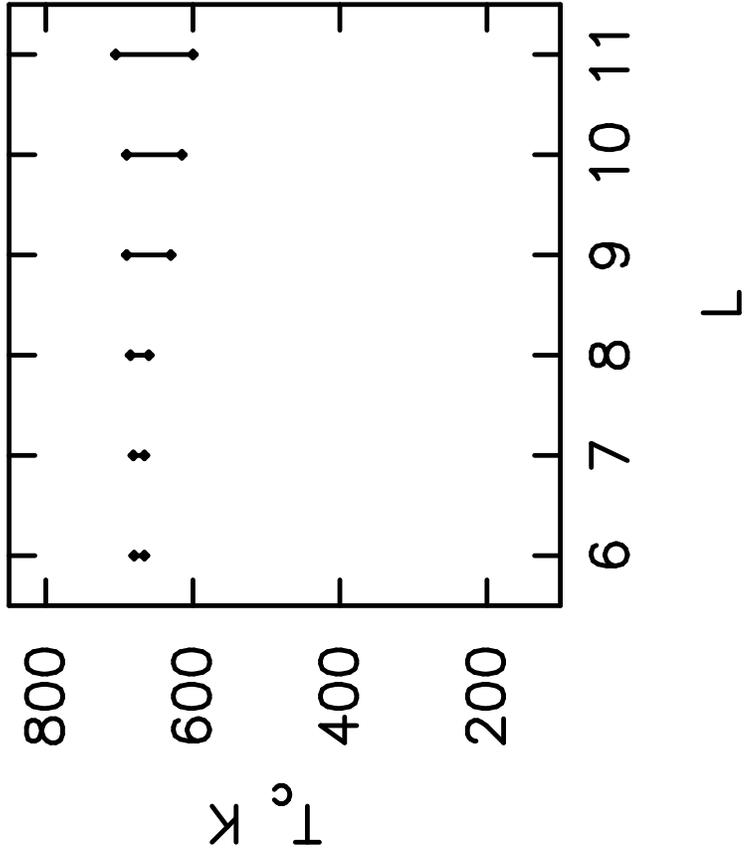
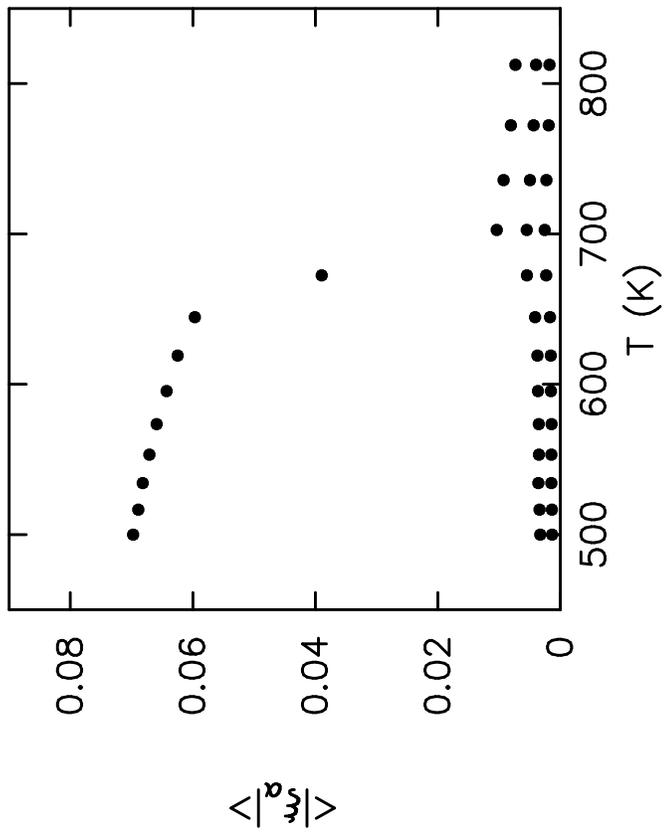
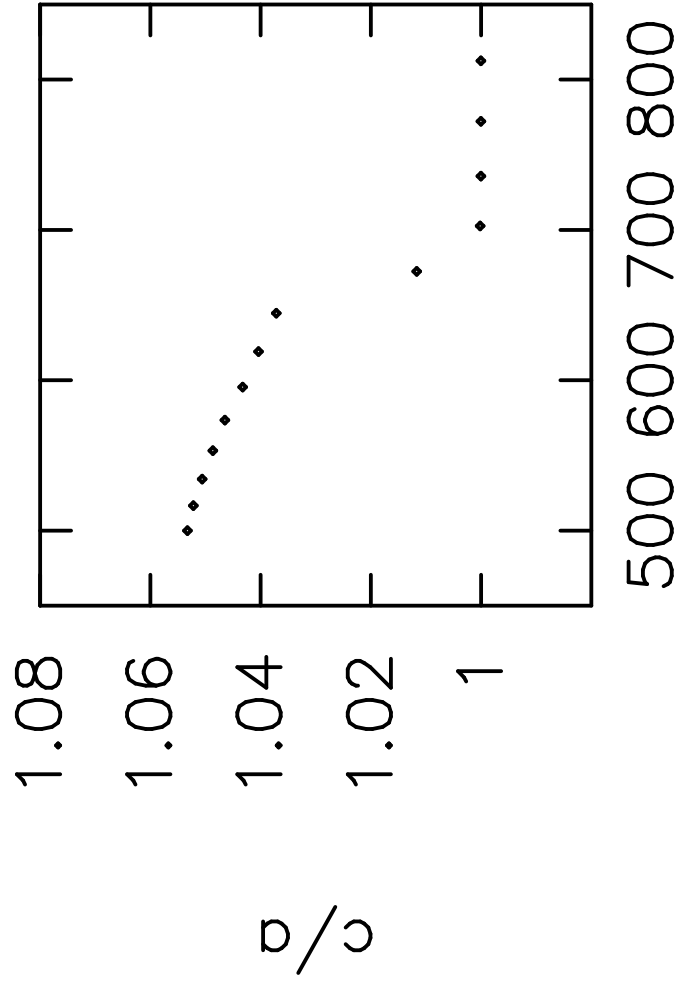
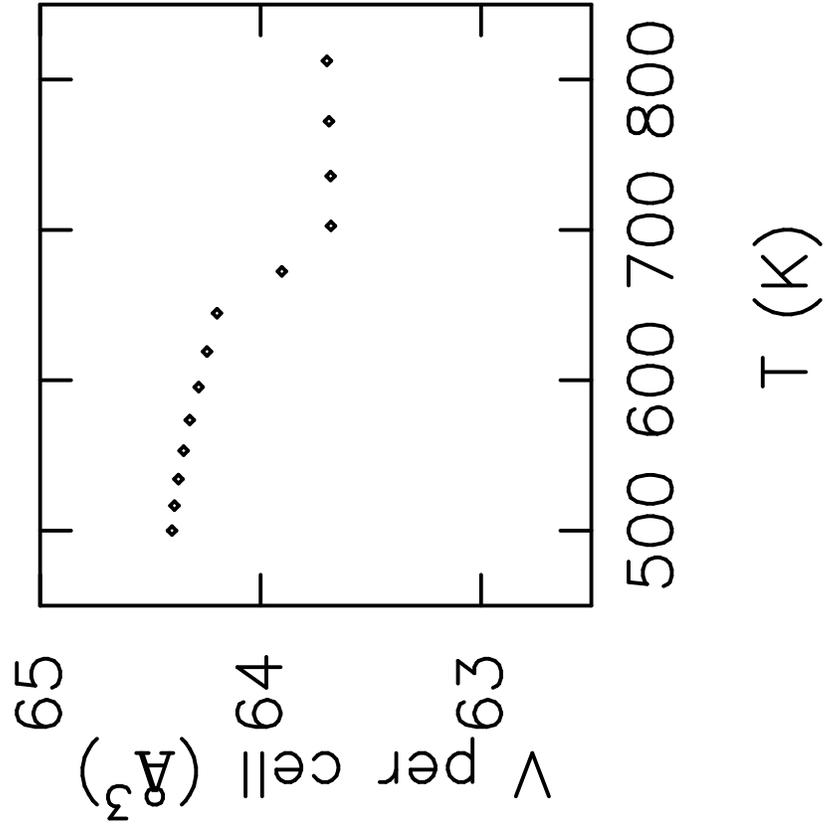
\end{document}